\documentclass[12pt]{article}

\begin{document}

\date{}
\title{Identical Quantum Particles and Weak Discernibility}
\author{Dennis Dieks and Marijn Versteegh \\ Institute for History and Foundations of Science \\
Utrecht University, P.O.Box 80.000 \\ 3508 TA Utrecht, The
Netherlands}
\maketitle

\begin{abstract}
We examine a recent argument that ``identical'' quantum particles
with an anti-symmetric state (fermions) are \textit{weakly
discernible} objects, just like irreflexively related ordinary
objects in situations with perfect symmetry (Black's spheres, for
example). We conclude that the argument uses a silent premise that
is not justified in the quantum case.
\end{abstract}

\section{Individuality}
Objects are individuals. One might think of their individuality as
something primitive, or as grounded in a metaphysical principle
not accessible to science---but here we shall not consider such
haecceistic conceptions. We take as our starting point that
entities must differ in their qualitative features, described by
the relevant scientific theory, in order to be treated as
different objects. This is in the spirit of Leibniz's principle of
the identity of indiscernibles (PII).

At first sight it may seem that PII cannot function as a general
basis for individuality, since it does not do justice to cases
with several objects in a symmetrical configuration. Think of Max
Black's spheres, of identical composition and two miles apart in a
relational space (\`{a} la Leibniz, not in Newtonian absolute
space where absolute positions could label the spheres); Kant's
enantiomorphic hands; or, for a mathematical example, the points
in the Euclidean plane. In all these cases there are obviously
more than one individual objects, but it also appears clear that
these objects share all their qualitative features. Both spheres
in Black's example have the same material properties and both are
at two miles distance from a sphere; similarly, Kant's hands have
the same internal geometric properties and are both mirror images
of a hand; and so on. However, the impression that PII is in
trouble here is superficial. As emphasized by Saunders
\cite{saundersleibn,saundersobj}, who takes his clue from Quine
\cite{quine}, it is essential to note that in these examples
\textit{irreflexive} qualitative relations exist between the
entities---relations an entity cannot have to itself. This
irreflexivity is sufficient to prove that PII is satisfied after
all.

There are some subtleties here that deserve attention. To discuss
these, let us first formalize the above argument. In formal
first-order languages we can \textit{define} identity ($=$):
\begin{equation}\label{id}
    s=t \equiv P(s) \leftrightarrow P(t),
\end{equation}
where $P$ denotes an arbitrary predicate in the language, and the
right-hand side of the definition stipulates that $s$ and $t$ can
replace each other, \textit{salva veritate}, in any $P$. This
definition captures PII, and the above discussion of
individuality, if the language is that of a physical theory, in
which the predicates refer to physical properties and relations
(and not to haecceities).

There can now be three kinds of discernibility
(\cite{saundersobj}). Two objects are \textit{absolutely
discernible} if there is a one-place predicate that applies to
only one of them; \textit{relatively discernible} if there is a
two-place predicate that applies to them in only one order; and
\textit{weakly discernible} if an \textit{irreflexive} two-place
predicate relates them. The latter possibility is relevant for our
examples. If there is an irreflexive but symmetric two-place
predicate $P(.,.)$ satisfied by $s$ and $t$, the definition
(\ref{id}) requires that if $s$ and $t$ are to be identical, we
must have:
\begin{equation}\label{}
    \forall x (P(s,x) \leftrightarrow P(t,x)).
\end{equation}
But this is false: in a valuation in which $P(s,t)$ is true,
$P(t,t)$ cannot be satisfied by virtue of the fact that $P$ is
irreflexive. Therefore, it immediately follows that PII is
satisfied by any two objects that are irreflexively related. The
only thing needed to dispel the impression that PII is in trouble
in these cases is therefore to apply the principle not only to
monadic predicates, but also to relations.

There might seem to be a remaining difficulty. In the above
argument we made use of the notion of a \textit{valuation}. A
valuation results from letting the names and bound variables in
the formulas of the language refer to specific elements of the
intended domain. In other words, in order to construct a
valuation, we have to name and distinguish the things we are
discussing. But this is an impossible task in the symmetrical
configurations we have been considering. Because of the symmetry,
any feature that can be attributed to one element of the domain
can also be attributed to any other. We can therefore not uniquely
refer and assign names on the basis of the given structure of
properties and relations. It is clearly impossible, for example,
to single out any \textit{specific} point in the Euclidean plane
on the basis of the properties of the plane and its points, even
if we include all relational properties. This impossibility might
appear to take the edge off the above argument; and it might be
thought to threaten the individuality of our entities after all
(because individuals can bear names, whereas no names can be given
here) \cite{keranen}.

The difficulty is only apparent, however. In order for the notions
of number and individuality to apply to the members of a domain it
is sufficient that a function \textit{exists} that maps the domain
one-to-one onto a set of labels, e.g.\ the set $\{1,2,...,n\}$
\cite[p.\ 457]{vanfraassen}. It is not needed that we can actually
\textit{construct} such a labelling. In all the cases we have been
considering the required mappings exist, and provide us with
valuations.

Although the name-giving problem is thus dissolved, it has drawn
our attention to an essential aspect of our examples of weak
discernibility: in all these cases the domains possessed a
structure that underwrote the existence of bijections to sets of
labels (e.g., sets of one or more natural numbers). As we shall
illustrate in a moment, the mere possibility of speaking about a
domain in terms of irreflexive relations is not enough to ensure
such an ``object structure'': it is possible to use properties or
relations talk even in situations in which there are no objects at
all. In such cases arguments about irreflexivity and weakly
discernible individuals obviously cannot come off the ground. This
means that we have to decide whether the concept of an object
applies in the first place, before we can use irreflexivity of the
relations to prove that the objects in question are weakly
discernible individuals.

This does not necessarily imply that there must be a division of
labor between ``objecthood providers'' and the irreflexive
relations. The relational structure may be the only structure of
the domain that is available and may provide our only access to
objecthood. The relations in our two physical examples---being at
a spatial distance from each other, being each other's mirror
image---indeed give us information about the presence of objects.
They are defined for relata that are physical things: things that
can be displaced with respect to each other, that can be reflected
and whose orientations can be compared. Such relations apply to
objects; conversely, if such relations apply, there are one or
more objects as relata. Objects are able to enter into more
relations, and this provides us with an additional test for
objecthood that applies to the cases mentioned thus far. Indeed,
why are we so sure that there are two Blackian spheres and two
Kantian hands? Our mind's eye sees Black's spheres at different
distances, and Kant's hands with different orientations, before
us. Speaking generally, when we break the symmetry of a
configuration of objects by introducing a reference object, gauge
or standard, the objects will become distinguishable with respect
to this standard. Thus we can name Black's spheres via their
unequal distances to a reference sphere and in Kant's universe we
may imagine a reference hand conventionally called ``left''.
Another example, relevant for our subsequent discussion of quantum
objects, is furnished by two oppositely directed arrows in an
otherwise empty Leibnizean world. If we conventionally fix a
standard of being ``up'', we break the symmetry and the individual
arrows become up or down---which verifies that they were weakly
discernible \textit{objects} in the original setting.

As already announced, it is essential for our subsequent argument
that irreflexive relations do not automatically connect objects,
and do not always pass the just-mentioned test. Consider the
example of Euros in a bank account (not coins in a piggy bank, but
transferable money in a real bank account). Imagine a situation in
which by virtue of some financial regulation the Euros in a
particular account can only be transferred to different one-Euro
accounts. So, in a complete money transfer an account with five
Euros, say, would be emptied and five different one-Euro accounts
would result. In this case the Euros in the original account stand
in an irreflexive relation to each other, namely ``only
transferable to different accounts''. But this does not make them
into physical objects. Analogously to the earlier cases, we could
try to exploit the irreflexive relations by introducing an
additional standard Euro that can only be transferred to one
specific account; or more directly, we may attempt to label the
Euros by means of the accounts they can end up in. However, this
does not achieve anything for the purpose of distinguishing
between the Euros in the account they are actually in. The
essential point is that the relations here do not pertain to
occurrent, actual, physical features of the situation; they do not
connect actual physical relata. Rather, the relational structure
is defined with respect to what \textit{would} result if the
actual situation \textit{were} \textit{changed}. There is of
course no doubt that five different one-Euro accounts (with
different account numbers) are five individuals; but this does not
mean that it makes sense to consider the five Euros as individual
objects before the transfer, when they are still in one common
account. On the contrary, the case of more than one money units in
one bank account is the standard example of classical
non-individuals, where only the account itself, with the total
amount of money in it, can be treated as possessing individuality
\cite{schrodinger,teller}. Although we may use relations and
things \textit{talk} here, there is nothing in the actual physical
situation that directly corresponds to this. Speaking of several
Euros in a bank account is a \textit{fa\c{c}on de parler} and does
not refer to actual physical objects. According to our best
theoretical understanding of the situation, statements like ``all
Euros in this account have the same value---namely one Euro'' are
not about actual physical things.

Summing up, we have found an important silent presupposition in
the argument for PII-based individuality in the presence of
irreflexive relations. Such relations can only be trusted to be
significant for the individuality issue if they are of the sort to
connect actual relata, objects. One way of verifying this is to
look into the relations' meaning and role in the pertinent
scientific theory. Another is to make use of a consequence of
objecthood: breaking of the perfect symmetry will result in the
possibility of giving names. This leads to a necessary condition,
a test: break the symmetry and see if names become assignable. The
breaking of the symmetry is analogous to the introduction of a
coordinates origin in describing a figure in plane geometry. If a
mapping to natural numbers exists in the presence of such a
reference point, it will still be there when the reference point
has been removed; what changes is the \textit{constructibility} of
the mapping. For the \textit{existence} of a mapping it is only
relevant whether the relations are such that they link actual
things; the introduction of an external vantage point makes it
possible to assign names, but does not change objecthood or the
number of objects.

\section{The Quantum Case}

One notorious interpretational problem of quantum theory concerns
so-called identical particles: particles of the same kind, like
electrons, protons or neutrons. It is a principle of quantum
theory that the state of a collection of such particles is
completely symmetric (in the case of bosons) or anti-symmetric
(fermions). This symmetrization postulate implies that all
one-particle states occur symmetrically in the total state of a
collection of identical particles. It follows that any property or
relation that can be attributed, on the basis of the total quantum
state, to one particle is attributable to all others as well.

The standard response to this conclusion is to say that identical
quantum particles lack individuality. This is tantamount to saying
that as far as the physical description goes they are no particles
at all: there may be ``many of them'', but this is like many Euros
in a bank account. It is better, according to this received view,
to renounce talk that suggests the existence of individual
particles---we should reconceptualize the situation in terms of
the excited states of a field (analogous to thinking of the Euros
in an account as one sum of money).

However, the situation also reminds us of the symmetric classical
cases described in the previous section. As we have seen there,
symmetry is not enough to prove the absence of non-haecceistic
individuality: particles may still be weakly discernible
individuals. Could it not be that in the quantum case there are
irreflexive physical relations between particles that guarantee
their individuality in the same way as they did for Black's
spheres, Kant's hands and Euclid's points? This is the position
adopted by Saunders, at least for the case of fermions
\cite{saundersobj}. Indeed, the anti-symmetry of the state of
many-fermions systems implies the existence of irreflexive
relations: intuitively speaking, the fermions in any pair stand in
the relation of ``occupying different one-particle states'', even
though any specific state description applicable to one of them
applies equally to the other. It is true (and noted by Saunders)
that for bosons with their symmetrical states this manoeuvre is
not available, so that collections of identical bosons are still
best understood as one whole. But the conclusion that standard
quantum mechanics\footnote{It is simple to promote quantum
particles to the status of unproblematic individuals by
\textit{adding} individuating characteristics to the descriptions
given by the theory, as is done, e.g., in the Bohm interpretation.
The problem is that these added features are empirically
superfluous---anyway, here we look only at standard quantum
mechanics, like Saunders does.} entails that fermions (these are
the ordinary matter particles; bosons are quanta of interaction
fields) are ordinary individuals---though only weakly
indiscernible---is surprising and highly significant by itself.

The technical details of the argument can be illustrated by the
example of two fermions in the singlet state. If $|\!\!\uparrow
\rangle $ and $|\!\!\downarrow \rangle $ stand for states with
spins directed upwards and downwards in a particular direction,
respectively, the anti-symmetrization principle requires that a
typical two-fermion state looks like
\begin{equation}\label{twofermions}
\frac{1}{\sqrt{2}}\{|\!\uparrow \rangle_1 |\downarrow \rangle_2 -
|\!\downarrow \rangle_1 |\uparrow \rangle_2 \},
\end{equation}
in which the subscripts $1$ and $2$ refer to the one-particle
state-spaces of which the total state-space (a Hilbert space) is
the tensor product. These two one-particle state-spaces are the
available candidates for the description of single particles, so
their labels are candidate names for the individual fermions (we
are assuming hypothetically that this notion makes sense). Now,
the anti-symmetry of the total state implies that the state
restricted to state-space $1$ is the same as the restricted state
defined in state-space $2$. (The ``partial traces'' are
$\frac{1}{2}\{ |\!\uparrow \rangle \langle \uparrow \!\!|  +
|\!\downarrow \rangle \langle \downarrow \!\! |\}$ in both cases.)
The \textit{total spin} has the definite value $0$ in state
(\ref{twofermions}); that is, state (\ref{twofermions}) is an
eigenstate of the operator $S_1 \otimes I + I \otimes S_2$.
Therefore, it seems natural to say that the two spins are
\textit{oppositely directed}. On the other hand, we cannot assign
a definite spin direction to the single particles because the up
and down states occur completely symmetrically in each of the
Hilbert spaces $1$ and $2$, respectively.

This situation is reminiscent of that of weakly discernible
classical objects; in particular it appears essentially the same
as the case of the two arrows. As we have seen in the latter case,
it is not possible to designate one of the arrows as up and the
other as down---but nevertheless there must be two individual
arrows in view of the oppositeness of their directions. Similarly,
in the fermion case with total spin zero we appear to have two
individual quantum objects with opposite spins.

\section{Quantum Individuals?}

On closer examination the similarity starts to fade away, however.
One should already become wary by the observation that the
irreflexive relations in the quantum case have a formal
representation that is quite different from that of their
classical counterparts in the cases we contemplated earlier. There
the relations could be formalized by ordinary predicates that can
be expressed as \textit{functions} of occurrent properties of the
individual objects (like ``up'' and ``down'', or $+1$ and $-1$,
with the correlation expressed by the fact that the sum of these
two quantities vanishes). By contrast, here we have that the total
system is in an \textit{eigenstate} of a linear \textit{operator}
in the total system's Hilbert space. Concomitant with this formal
difference is an essential difference in interpretation: according
to standard quantum mechanics, quantum states should be
interpreted in terms of possible \textit{measurement results} and
their probabilities, rather than in terms of occurrent properties.
In the case at hand, a system in an eigenstate of the total spin
operator with eigenvalue $0$, this means that a measurement of the
total spin will have the outcome $0$ with probability $1$. In this
special case, in which the outcome is certain (probability $1$),
it is plausible and harmless to think that the total system
possesses the property ``total spin $0$'' also independently of
measurement; but even if we assume this, this total spin cannot be
understood as being composed of definite spin values of the two
subsystems. Although it is of course possible to perform
individual spin \textit{measurements} on the subsystems (whose
possible outcomes and corresponding probabilities are predictable
from the total quantum state), there are no corresponding
\textit{occurrent spin properties} in the subsystems,
independently of measurement. In the singlet state
(\ref{twofermions}) the prediction of quantum mechanics is that
individual spin \textit{measurements} will with certainty yield
opposite results, summing up to $0$; but on the pain of running
into paradoxes and no-go theorems it cannot be maintained that
these results reveal oppositely directed spins that already were
there before the measurements. This is an example of the notorious
``holism'' of quantum mechanics: definite properties of a
composite system need not be reducible to properties of its parts.

This suggests that the correct analogue to the quantum case is not
provided by two oppositely directed classical arrows, but rather
by a two-Euro account that can be \textit{transformed} (upon
``measurement'', i.e.\ the \textit{intervention} brought about by
a money transfer) into two distinct one-Euro-accounts. To
investigate further whether or not the quantum relations connect
actual physical relata, we may copy the strategy followed in the
classical case, namely breaking the symmetry and seeing whether in
the resulting situation the quantum relations can serve as
name-givers. This cannot work as long as we stay within a
many-fermions system: quantum mechanics forbids fermion systems
that are not in an anti-symmetric state---accordingly, it is a
matter of lawlike principle that the only relations fermions can
possess with respect to each other are perfectly symmetrical. So
if we want to break the symmetry this should be done by the
introduction of a standard that is external to the fermion system.
After this we are in a position to verify whether the quantum
relations with respect to the introduced standard make the
individual fermions distinguishable.

To see the inevitability of a negative outcome of any such
attempt, consider an arbitrary system of identical quantum
particles to which a gauge system has been added without any
disturbance (i.e., the total state is the product of the original
symmetrical or anti-symmetrical identical particles state and the
state of the gauge system). Let the new total state be denoted by
$|\Psi\rangle$. Any quantum relation in this state between the
gauge system and one of the identical particles, described in
subspace $j$, say, has the form $\langle\Psi|A(g,j)|\Psi\rangle$.
Here $A(g,j)$ is a hermitian operator working in the state-spaces
of the gauge system $g$ and identical particle $j$. We can now use
the (anti)-symmetry of the original identical particles state to
show that the gauge system stands in exactly the same relations to
\textit{all} identical particles. The (anti)-symmetry entails that
$P_{ij}|\Psi\rangle = \pm |\Psi\rangle$, where $P_{ij}$ stands for
the operator that permutes indices $i$ and $j$. Now,
\begin{eqnarray*}
\langle\Psi|A(g,j)|\Psi\rangle =
\langle P_{ij}\Psi|A(g,j)|P_{ij}\Psi\rangle = \\
\langle\Psi|P^{-1}_{ij}A(g,j)P_{ij}|\Psi\rangle =
\langle\Psi|A(g,i)|\Psi\rangle.
\end{eqnarray*}
In other words, any quantum relation the gauge system has to $j$,
it also has to $i$, for arbitrary values of $i$ and $j$. That
means that these quantum relations have no discriminating value in
the situation as it actually is, without measurement interventions
and the disturbances caused by them.

It must be stressed that if the situation \textit{is}
\textit{changed} by a measurement interaction, distinct individual
results may arise\footnote{The expressions
$\langle\Psi|A(g,j)|\Psi\rangle$ are in this case interpreted as
\textit{expectation values}, averages over very many experimental
trials.}, just like in the case of the opposite spin results in
measurements in the total spin $0$ state (\ref{twofermions}). But
we are here interested in the question of whether a many-fermions
system \textit{as it is} can be regarded as a collection of weakly
discernible individuals; not in the question of whether such a
system can be \textit{transformed} into a collection of
individuals. The fermions behave like money units in a bank
account: it does not matter what external standard we introduce,
it will always possess the same relations to all (hypothetically
present) entities. This leaves us without evidence that there
\textit{are} any actual objects composing many-fermions systems.
This stands in sharp contrast to our earlier cases of classical
weakly discernible objects.

\section{Conclusion}

There is an essential difference between quantum mechanical
many-fermion systems and classical collections of weakly
discernible objects. In the latter case the objects are nameable
\textit{in abstracto}, although the symmetry of the situation
makes it impossible to actually assign names. This is a strange
situation; but the air of paradox is dispelled when we apply the
concept of weak discernibility. The strangeness of the quantum
case runs much deeper. There is no sign within standard quantum
mechanics that identical fermions are things at all; the
irreflexivity of relations does not help us here. Quantum
relations have an interpretation not in terms of what is actual,
but rather via what \textit{could} happen in case of a
measurement; and they cannot be used in a name-giving procedure
after the introduction of an external standard. There is therefore
no evidence that the quantum relations between fermions connect
any actual physical objects. As far as standard quantum mechanics
goes, identical fermions are not discernible, not even weakly.
Conventional wisdom, saying that systems of identical quantum
particles should be considered as one whole, like an amount of
money in a bank account, appears to be right after all.

\end{document}